\documentclass[12pt,a4paper]{article}
\usepackage{graphicx}
\pdfoutput=1

\title{\bf A Monte Carlo study of double logarithms in the small x region} 
\author{G. Chachamis, A. Sabio Vera\\ \\
{\small Instituto de F{\' \i}sica Te{\' o}rica UAM/CSIC, Nicol{\'a}s Cabrera 15}\\ 
{\small \& Universidad Aut{\' o}noma de Madrid, E-28049 Madrid, Spain.}
}

\begin{document} 

\maketitle 

\abstract

We investigate the effect of the resummation of collinear double logarithms in the BFKL gluon Green function using the Monte Carlo event generator {\tt BFKLex}. The resummed collinear terms in transverse momentum space were calculated in Ref.~\cite{Vera:2005jt} and correspond to the addition to the NLO BFKL kernel of a Bessel function of the first kind whose argument contains the strong coupling and a double logarithm of the ratio of the squared transverse momenta of the reggeized gluons. We discuss how these additional terms improve the collinear convergence of the whole approach and reduce the asymptotic growth with energy of cross sections. Taking advantage of the Monte Carlo implementation, we show how the new results reduce the diffusion of the gluon ladder into infrared and ultraviolet transverse momentum scales, while strongly affecting final state configurations by reducing the 
mini-jet multiplicity.

\section{Brief introduction}

A fascinating area of study within strong interactions at high energies is the Balitsky-Fadin-Kuraev-Lipatov (BFKL) approach. The key idea in this formalism is that, when the center-of-mass energy $\sqrt{s} \to \infty$, terms of the form {$\alpha_s^n 
\log^n{\left(s\right)} \sim \alpha_s^n \left(y_A-y_B\right)^n$} (where $y_{A,B}$ are the rapidities of tagged particles in the final state) must be resummed in order to accurately describe experimental observables. In this limit there is decoupling between transverse and longitudinal degrees of freedom which allows to evaluate cross sections in the factorized form: 
\begin{eqnarray}
\sigma^{\rm LL} &=& \sum_{n=0}^\infty {\cal C}_n^{\rm LL}  \alpha_s^n 
\int_{y_B}^{y_A} d y_1 \int_{y_B}^{y_1} d y_2 \dots \int_{y_B}^{y_{n-1}} d y_n \nonumber\\ 
&=& \sum_{n=0}^\infty \frac{{\cal C}_n^{\rm LL}}{n!} 
\underbrace{\alpha_s^n \left(y_A-y_B\right)^n }_{\rm LL} \nonumber
\end{eqnarray}
where LL stands for the  leading log approximation and $y_i$ correspond to the rapidity of emitted particles. The LL BFKL formalism allows us to calculate the coefficients ${\cal C}_n^{\rm LL}$~\cite{Lipatov:1985uk,Balitsky:1978ic,Kuraev:1977fs,Kuraev:1976ge,Lipatov:1976zz,Fadin:1975cb}. The next-to-leading log approximation (NLL)~\cite{Fadin:1998py,Ciafaloni:1998gs} is more complicated since it is sensitive to the running of the strong coupling and  to the choice of energy scale in the logarithms~\cite{Forshaw:2000hv,Chachamis:2004ab,Forshaw:1999xm,Schmidt:1999mz}. We can parametrize the freedom in the choice of these two scales, respectively, by introducing the constants 
${\cal A}$ and ${\cal B}$ in the previous expression: 
\begin{eqnarray}
\sigma^{LL+NLL} &=& 
\sum_{n=1}^\infty \frac{{\cal C}_n^{\rm LL} }{n!}  \left(\alpha_s- {\cal A} \alpha_s^2\right)^n \left(y_A-y_B - {\cal B}\right)^n \nonumber\\
&&\hspace{-2.4cm}= \sigma^{\rm LL}  - \sum_{n=1}^\infty   \frac{\left({\cal B}  \, {\cal C}_n^{\rm LL} +  (n-1) \, {\cal A} 
\, {\cal C}_{n-1}^{\rm LL} \right)}{(n-1)!}  \underbrace{ \alpha_s^n 
\left(y_A-y_B\right)^{n-1}}_{\rm NLL} + \dots \nonumber
\end{eqnarray}
We see that at NLL a power in $\log{s}$ is lost w.r.t. the power of the coupling. In this formalism we can then calculate cross sections using the factorization formula (with $Y\simeq \ln{s}$)
 \begin{eqnarray}
\sigma (Q_1,Q_2,Y) = \int d^2 \vec{k}_A d^2 \vec{k}_B \, \underbrace{\phi_A(Q_1,\vec{k}_A) \, 
\phi_B(Q_2,\vec{k}_B)}_{\rm PROCESS-DEPENDENT} \, \underbrace{f (\vec{k}_A,\vec{k}_B,Y)}_{\rm UNIVERSAL}, \nonumber
\end{eqnarray}
where $\phi_{A,B}$ are process-dependent impact factors which are functions of some external scale, $Q_{1,2}$, and some internal momentum for reggeized gluons, $\vec{k}_{A,B}$.  The Green function $f$ is universal and depends on $\vec{k}_{A,B}$ and the energy of the process $\sim e^{Y/2}$. It corresponds to the solution of the BFKL equation which at LL can be written in iterative form~\cite{Schmidt:1996fg} in transverse momentum representation as
\begin{eqnarray}
f &=& e^{\omega \left(\vec{k}_A\right) Y}  \Bigg\{\delta^{(2)} \left(\vec{k}_A-\vec{k}_B\right) + \sum_{n=1}^\infty \prod_{i=1}^n \frac{\alpha_s N_c}{\pi}  \int d^2 \vec{k}_i  
\frac{\theta\left(k_i^2-\lambda^2\right)}{\pi k_i^2} \nonumber\\
&&\hspace{-1.2cm}\int_0^{y_{i-1}} \hspace{-.3cm}d y_i e^{\left(\omega \left(\vec{k}_A+\sum_{l=1}^i \vec{k}_l\right) -\omega \left(\vec{k}_A+\sum_{l=1}^{i-1} \vec{k}_l\right)\right) y_i} \delta^{(2)} 
\left(\vec{k}_A+ \sum_{l=1}^n \vec{k}_l - \vec{k}_B\right)\Bigg\}, \nonumber
 \end{eqnarray}
where the gluon Regge trajectory reads
\begin{eqnarray}
\omega \left(\vec{q}\right) &=& - \frac{\alpha_s N_c}{\pi} \log{\frac{q^2}{\lambda^2}}. \nonumber
\end{eqnarray}
$\lambda$ is a regulator of infrared divergencies. This solution has been studied at length in a series of papers. It serves as the basis to construct the Monte Carlo event generator {\tt BFKLex} which has had multiple 
applications in collider phenomenology and more formal studies~\cite{Chachamis:2013rca,Caporale:2013bva,Chachamis:2012qw,Chachamis:2012fk,Chachamis:2011nz,Chachamis:2011rw}. In the following Section we focus on a discussion of this iterative solution at NLL and, in particular, on those terms which drive its collinear behavior. 

\section{Resummation of the leading collinear contributions}

\begin{figure}
\includegraphics[height=10cm]{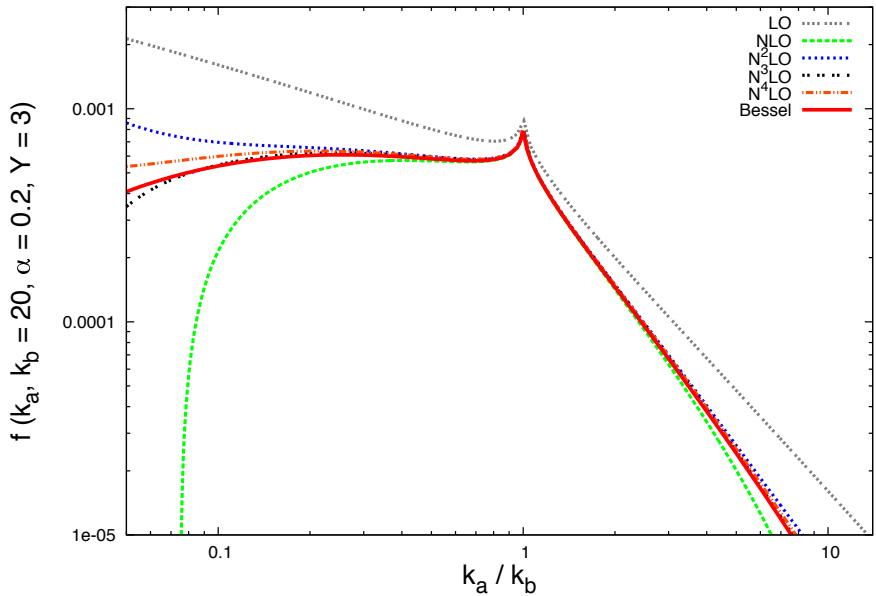}
\vspace{-1cm}
\caption{Collinear behavior of gluon Green function including collinear contributions}
\label{Coll-Bessel}
\end{figure}

At NLL there is a dominant term in the BFKL kernel in the collinear regions, which correspond to the limits $\vec{k}_A^2$ being much bigger or smaller than  $\vec{k}_B^2$, taking the form of a double log contribution. This term can be introduced in the previous iterative representation of the gluon Green function making the replacement
\begin{eqnarray}
\theta \left(k_i^2-\lambda^2\right) \to \theta \left(k_i^2-\lambda^2\right)  
- \underbrace{\frac{\bar{\alpha}_s}{4} \ln^2{\left(\frac{\vec{k}_A^2}{\left(\vec{k}_A+\vec{k}_i\right)^2}\right)}}_{\rm NLL}. 
\end{eqnarray}
This new contribution generates a large instability in the BFKL formalism when it is applied to cross sections  whose impact factors are not very narrowly localized in transverse momentum space. Since a typical impact factor allows for configurations where $\vec{k}_A^2$ can be quite different to $\vec{k}_B^2$ then this is an important problem which needs to be investigated in detail. The original BFKL formulation can be applied to processes where $\vec{k}_A^2 \simeq \vec{k}_B^2$, if we want to extend its applicability then it is needed to introduce all-orders collinear corrections to its kernel. This has been addressed in~\cite{Salam:1998tj,Ciafaloni:2003ek}. In~\cite{Vera:2005jt}  it was shown that the collinear corrections in transverse momentum space have the following structure when expanded in a perturbative series:
\begin{eqnarray}
\theta \left(k_i^2-\lambda^2\right) \to \theta \left(k_i^2-\lambda^2\right)  + \sum_{n=1}^\infty 
\frac{\left(-\bar{\alpha}_s\right)^n}{2^n n! (n+1)!} \ln^{2n}{\left(\frac{\vec{k}_A^2}{\left(\vec{k}_A+\vec{k}_i\right)^2}\right)}. 
\label{SumBessel}
\end{eqnarray}
In order to understand what it is the effect of this sum we have plotted the gluon Green function for a fixed coupling of $\bar{\alpha}_s=0.2$ and $Y=3$ with $k_b = 20$ versus $k_a$ in Fig.~\ref{Coll-Bessel} (all the transverse momenta in this paper are written in GeV). 
In this figure we compare the LL result in collinear and anti-collinear regions with the higher order corrections. We can add the terms in the sum of Eq.~(\ref{SumBessel}) one by one and observe that the converge is very good when we are not very far from the region $\vec{k}_A^2 \simeq \vec{k}_B^2$. We see that the higher order collinear corrections are needed to avoid a negative Green function (the NLO truncation goes negative rather quickly). The continuous line corresponds to the summation of an infinite number of terms in Eq.~(\ref{SumBessel}). In this case it was shown in~\cite{Vera:2005jt} that this resums to a Bessel function of the first kind (this approach, in $\gamma$-space, has already been successfully applied in many phenomenological applications~\cite{Vera:2006un,Vera:2007kn,Caporale:2007vs,Vera:2007dr,Caporale:2008fj,Hentschinski:2012kr,Hentschinski:2013id,Caporale:2013uva,Chachamis:2015ona}). A similar conclusion has recently been reached in coordinate space in~\cite{Iancu:2015vea}. 

What we have discussed in this short Section accounts for the leading collinear contributions. We will show next how to deal with a more complete set of them. 

\section{Bessel resummation in the {\tt BFKLex} Monte Carlo event generator}

We have implemented the collinear resummation in transverse momentum space in the {\tt BFKLex} Monte Carlo event generator including subleading logarithmic contributions and the full NLL BFKL kernel. The prescription was shown in~\cite{Vera:2005jt} and we explain it here for completeness. We keep the NLO BFKL kernel untouched and only add NNLO and beyond higher order corrections. To do so we  remove from the original NLO kernel the double log
\begin{eqnarray}
- \frac{\bar{\alpha}^2_s}{4} \frac{1}{(\vec{q}-\vec{k})^2}
\ln^2{\left(\frac{q^2}{k^2}\right)}
\label{singlelog}
\end{eqnarray}
and we replace it by the resummed expression
\begin{eqnarray}
 \Bigg\{\left(\frac{q^2}{k^2}\right)^{- b \, \bar{\alpha}_s
\frac{|k-q|}{k-q}} \sqrt{\frac{2(\bar{\alpha}_s + a \, \bar{\alpha}_s^2)}{\ln^2{\left(\frac{q^2}{k^2}\right)}}} J_1 \left(\sqrt{2(\bar{\alpha}_s + a \, \bar{\alpha}_s^2) \ln^2{\left(\frac{q^2}{k^2}\right)}}\right)\nonumber\\
&&\hspace{-9cm}- \bar{\alpha}_s - a \, \bar{\alpha}_s^2 + b \, \bar{\alpha}_s^2
\frac{|k-q|}{k-q} \ln{\left(\frac{q^2}{k^2}\right)} \Bigg\}\frac{1}{(\vec{q}-\vec{k})^2}
\end{eqnarray}
where $J_1$ is the Bessel function of the first kind and 
\begin{eqnarray}
a &=& \frac{5}{12} \frac{\beta_0}{N_c} - \frac{13}{36}\frac{n_f}{N_c^3} - \frac{55}{36};\\
b &=& - \frac{1}{8}  \frac{\beta_0}{N_c} - \frac{1}{6}\frac{n_f}{N_c^3} - \frac{11}{12};
\end{eqnarray}
take into account the subleading collinear logs (the previous section corresponds to the simple case with $a=b=0$, in this work we focus on the scale invariant part of the NLO kernel and took $\beta_0=0$ and $n_f=0$). Notice that the perturbative expansion of the Bessel function generates back, as the first term, the double log of Eq.~(\ref{singlelog}) which we had removed. This ensures that, up to NLO, we are in agreement with the results of Fadin and Lipatov. 

\begin{figure}
\includegraphics[height=10cm]{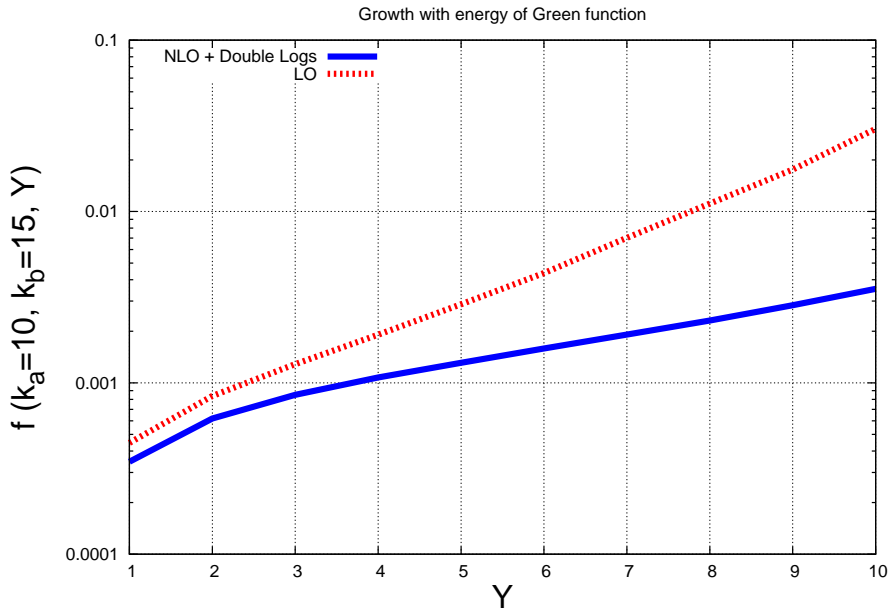}
\vspace{-1cm}
\caption{Growth with energy of the gluon Green function}
\label{Growth-Energy}
\end{figure}

\begin{figure}
\begin{center}
\includegraphics[height=10cm]{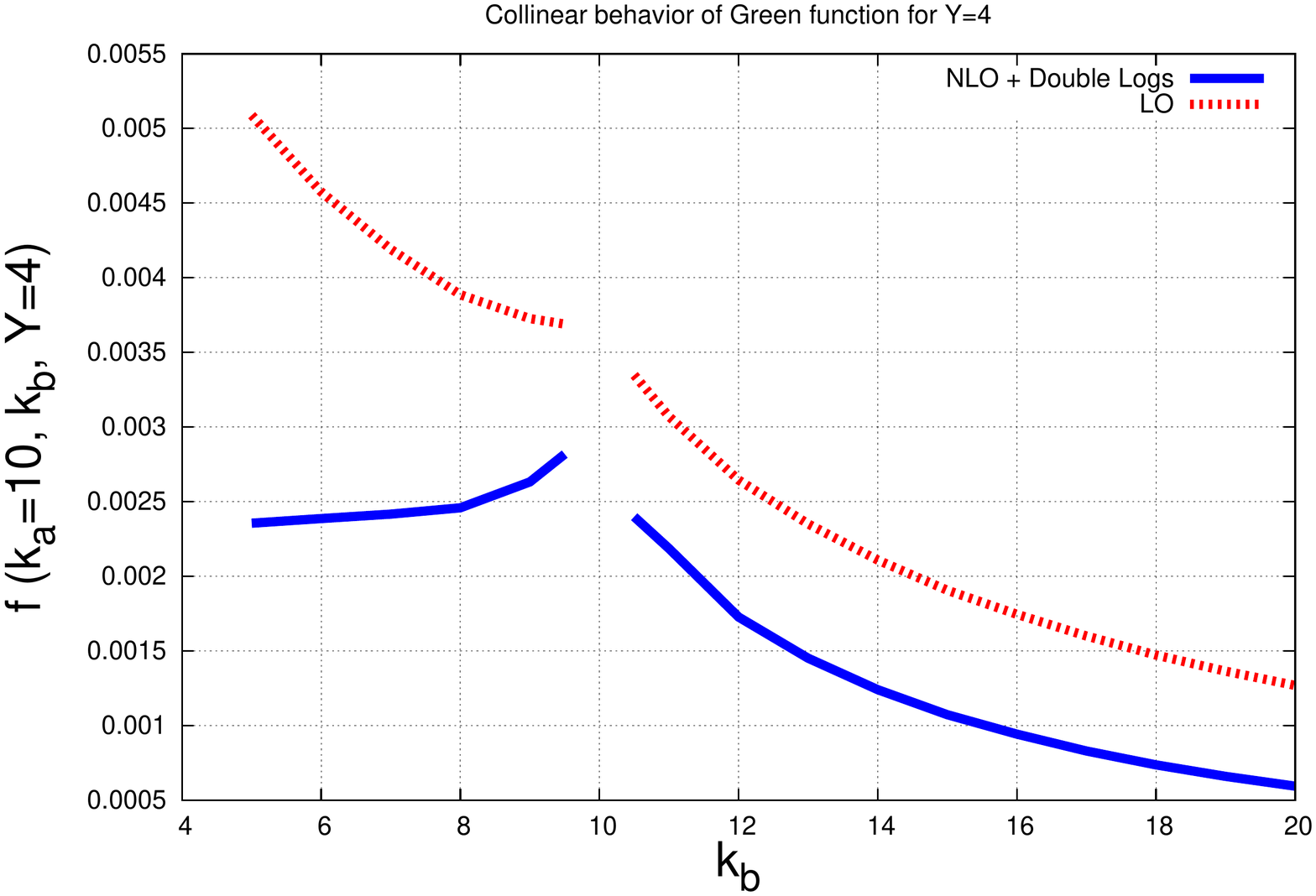}
\vspace{-1cm}\\
\includegraphics[height=10cm]{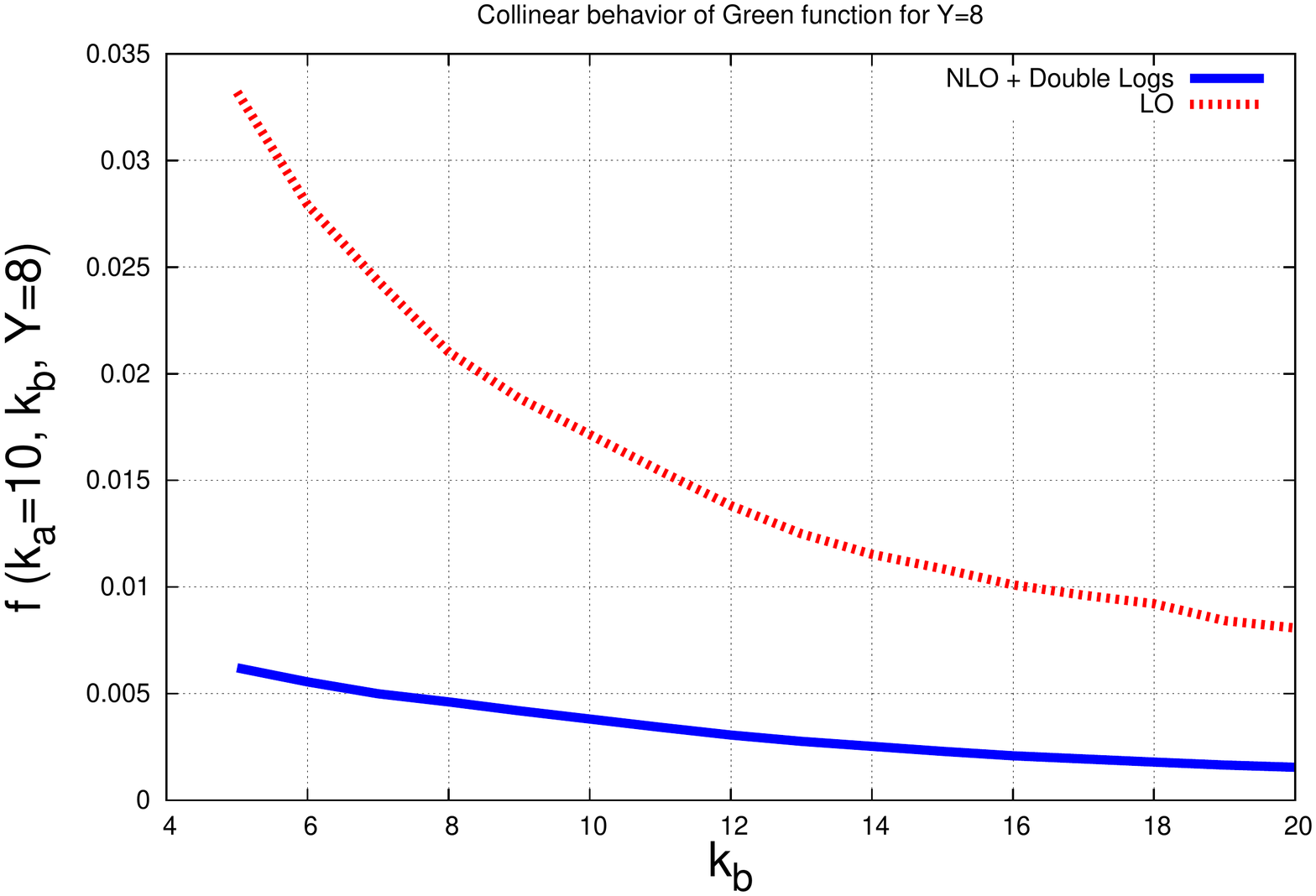}
\end{center}
\vspace{-1cm}
\caption{Collinear behavior of gluon Green function for different values of the scattering energy}
\label{Collinear-Y}
\end{figure}

The effect of this correction on the gluon Green function is quite dramatic. It greatly reduces its growth with energy as can be seen in Fig.~\ref{Growth-Energy}. 
This reduction is noticeable already at very low values of $Y$. The effect is also very large, as it should be, in the collinear regions which are largely modified, but positive, when the Bessel resummation is introduced, as can be seen in Fig.~\ref{Collinear-Y}, where we show the collinear behavior for two different energy values. 

In the following Section we discuss the effects of the double log resummation for a couple of more exclusive quantities: the so-called diffusion cigar and the multiplicity distributions. 

\section{The diffusion picture and multiplicities}

\begin{figure}
\hspace{-1.5cm}
\includegraphics[height=6.5cm]{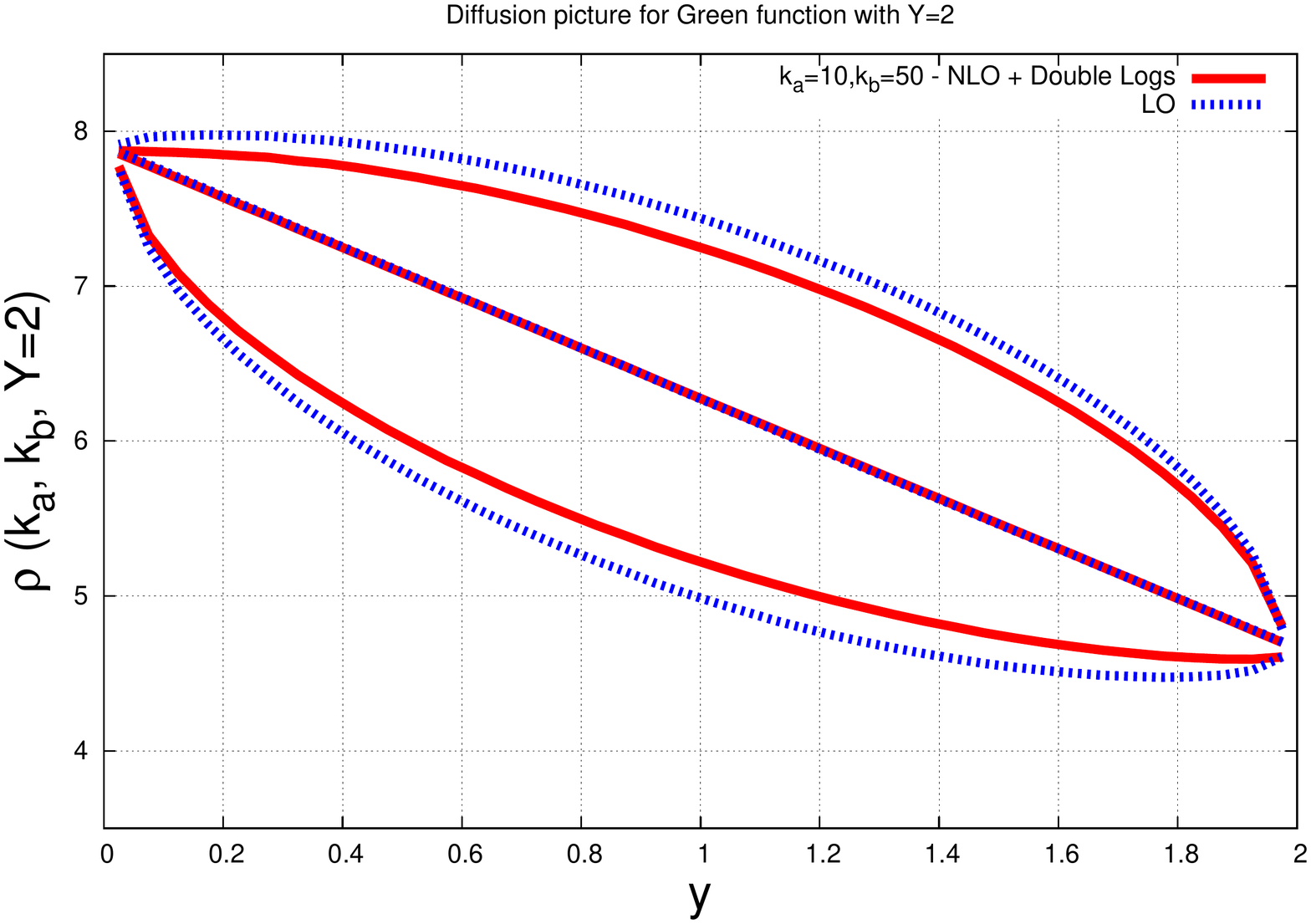}\includegraphics[height=6.5cm]{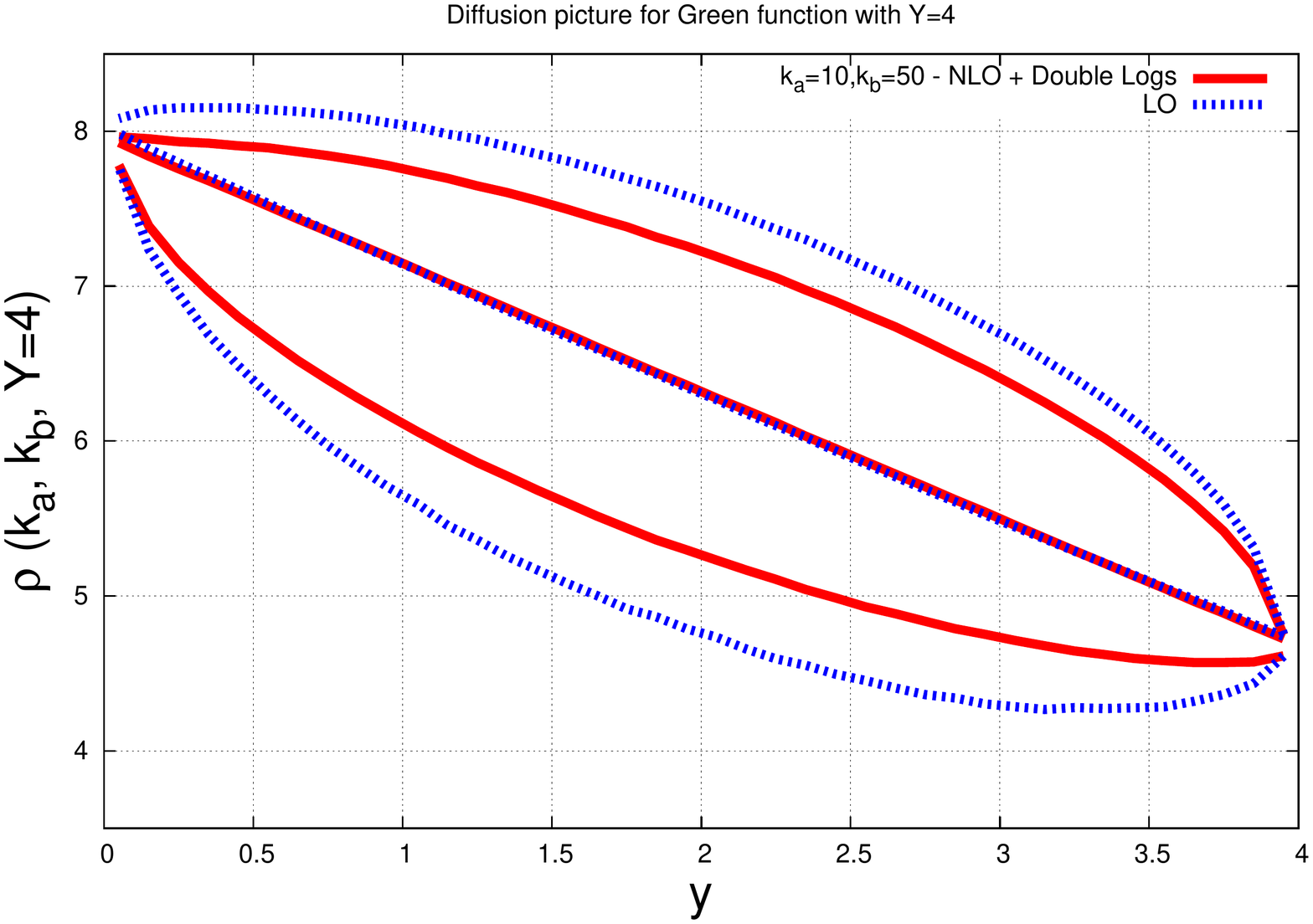}

\hspace{-1.5cm}\includegraphics[height=6.5cm]{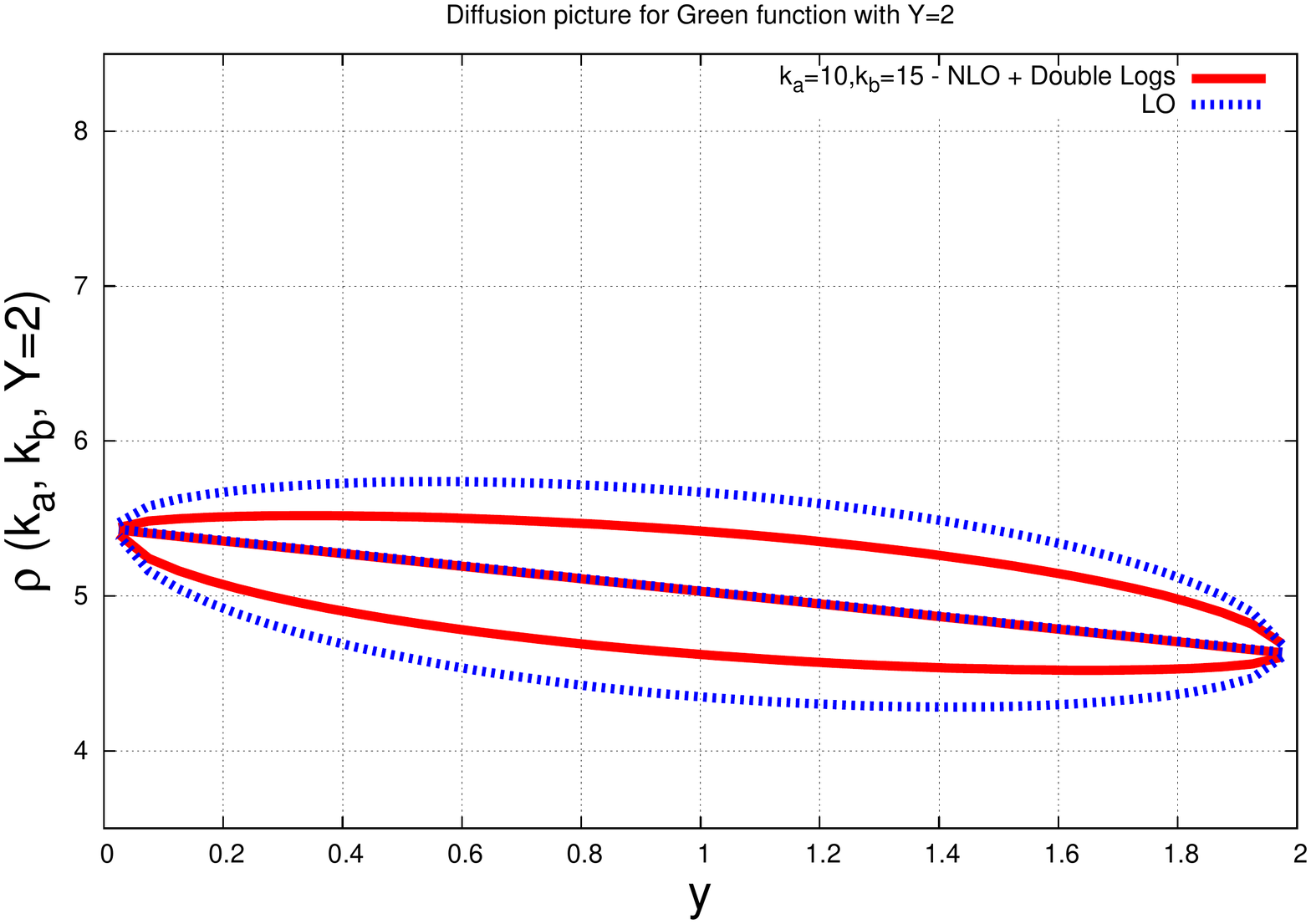}\includegraphics[height=6.5cm]{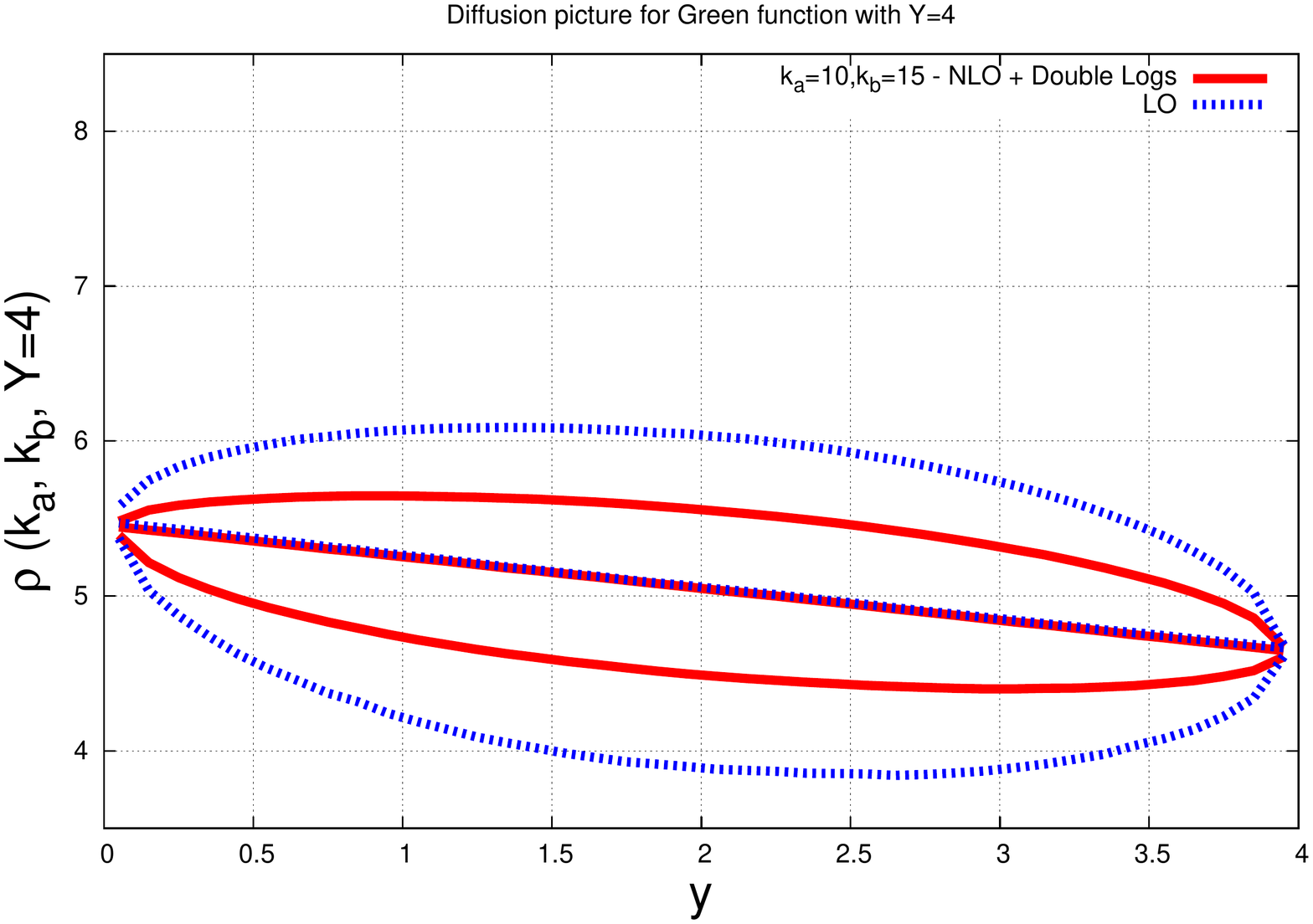}
\caption{Diffusion plots for different external transverse momenta configurations and different center-of-mass energies.}
\label{Melon-Y}
\end{figure}

An interesting feature arises when the collinear terms are included in our analysis: the typical transverse scales in the gluon ladder tend to be more localized around the external transverse scales and the diffusion towards infrared and ultraviolet regions is largely reduced. The quantity we call $\rho$ measures the mean value for the average transverse scale (the straight lines in Fig.~\ref{Melon-Y})  of those reggeized gluons propagating in a rapidity gap centered at a typical rapidity $y$. 
The curved lines in the same figure indicate the root mean square deviation from the mean value. We can see that, as it is well known, the diffusion or width of the cigar-shaped figures is broader as the total available energy, or total rapidity span increases (in the figure we chose two plots with $Y=2$ and $Y=4$, and note that $y$ varies from 0 to $Y$). In the upper and lower plots of the figure we chose asymmetric and symmetric external transverse scales, respectively. 

\begin{figure}
\hspace{-1.cm}\includegraphics[height=6.5cm]{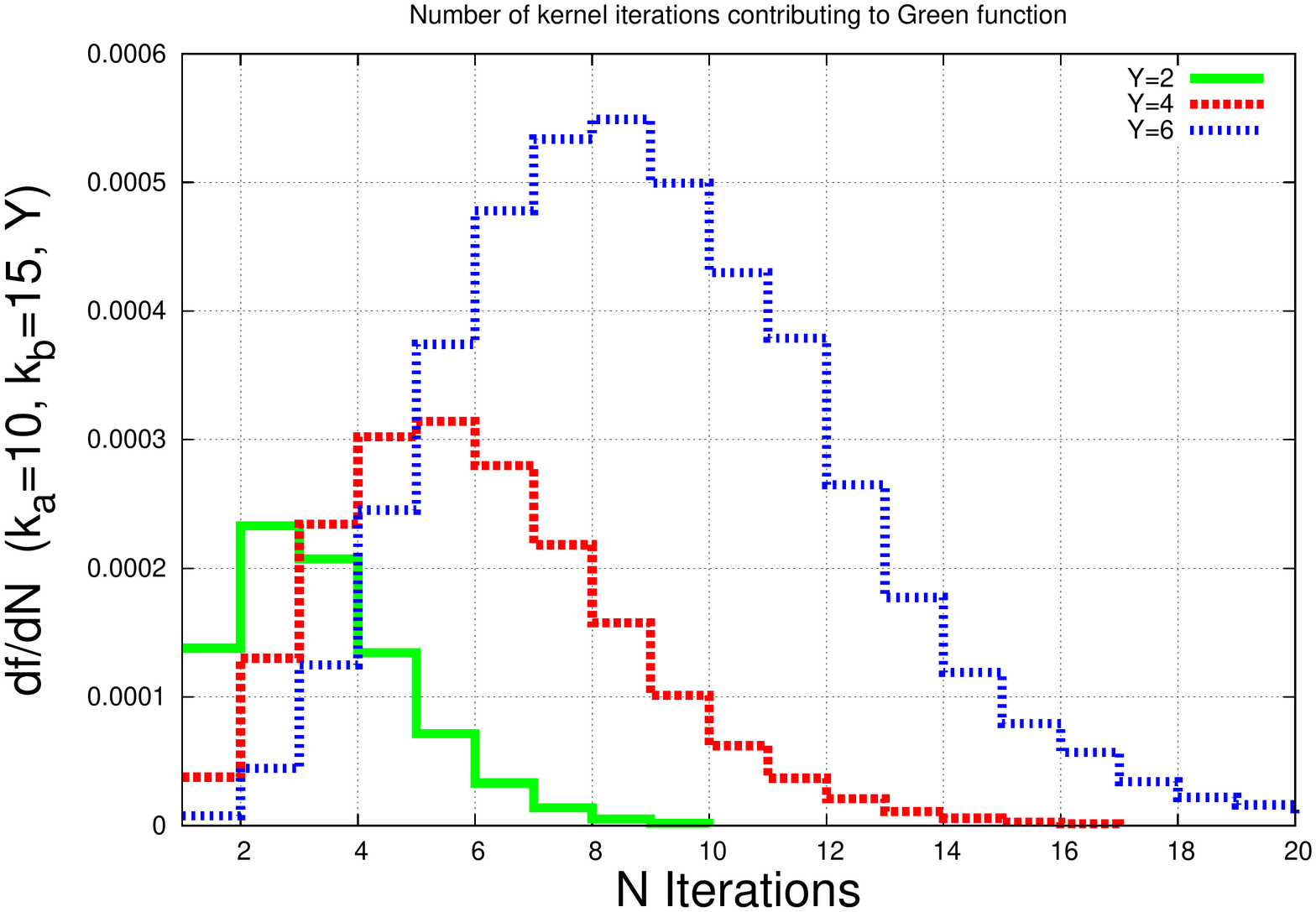}\includegraphics[height=6.5cm]{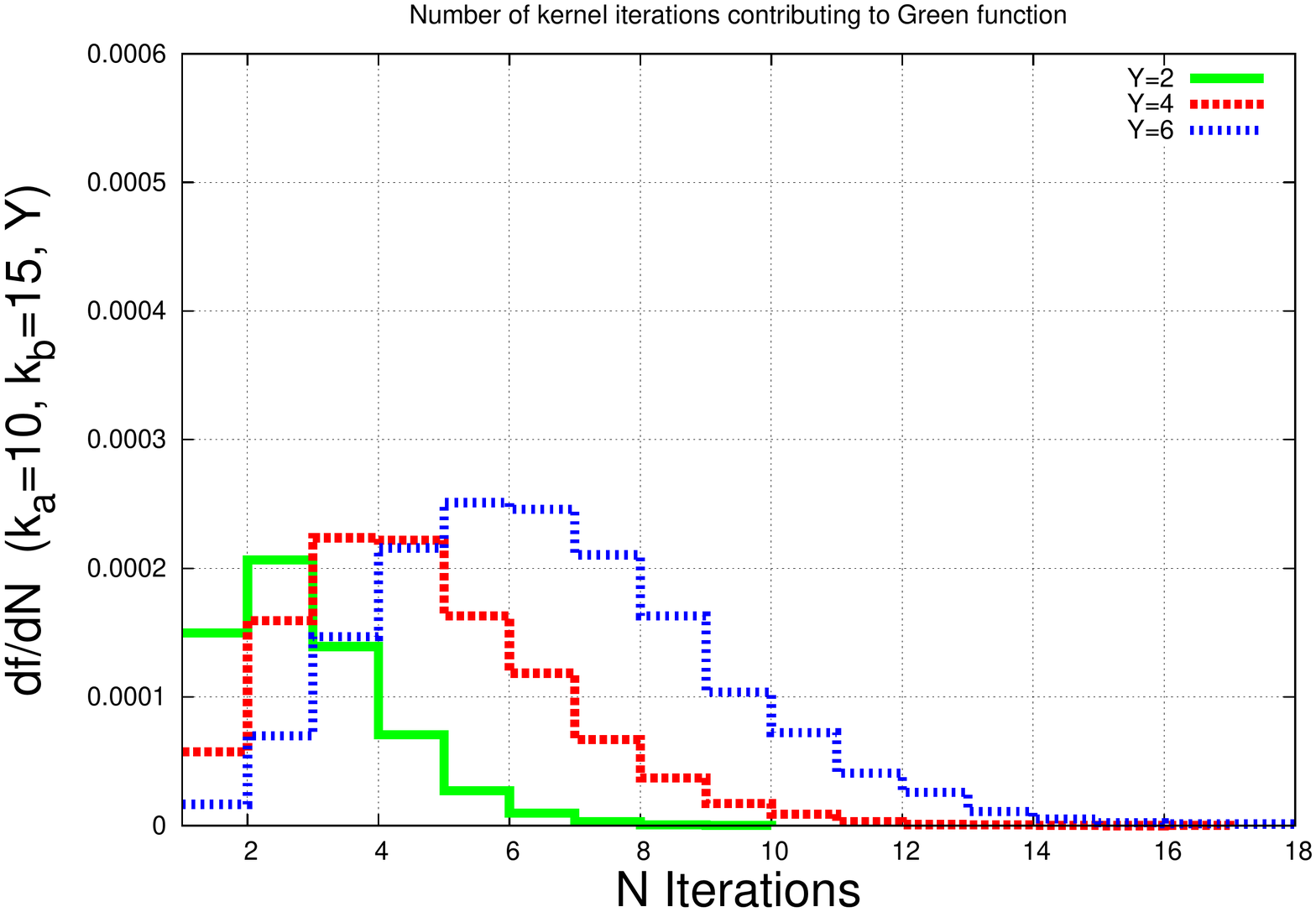}
\caption{Distribution in the number of iterations of the BFKL kernel needed to construct the gluon Green function.}
\label{Multiplicities}
\end{figure}

A striking new feature is revealed when looking at the number of iterations of the kernel needed to construct the solution to the BFKL equation. This is shown in Fig.~\ref{Multiplicities} al LL (left) and NLL +Bessel (right). The Green function at each $Y$ corresponds to the areas under the different curves there depicted.  In both cases we find Poissonian distributions in the number of iterations of the kernel, with a huge suppression in the case with collinear resummation (right plot). The remarkable fact is the position of the maxima in these distributions: in the collinearly-improved case we find that at larger rapidities these maxima are rather smaller than at LL. For rapidity $Y=4$ at LL it is placed at $N\simeq 5$ while with NLL+Bessel it lies at $N \simeq 3$. For $Y=6$ we have $N \simeq 8$ at LL versus 
$N\simeq 5$ in the NLL+Bessel case. The maximal $N$ with a significant contribution to the Green function also decreases a significant amount when moving from LL to a calculation with higher orders: at $Y=4,6$, $N_{\rm max}=16,22$, respectively, 
at LL while $N_{\rm max}=12,15$ at higher orders. 

This reduction of the effective final state multiplicity has phenomenological consequences which we will describe in a future work. We believe the implementation of higher order corrections into a Monte Carlo event generator can be a useful tool to test the high energy limit of QCD at present and future colliders. It is needed to define new, more exclusive, observables to find imprints of this interesting kinematical region.

\section{Summary \& Outlook}

We have presented a collinearly improved version of the NLL BFKL kernel which is valid in transverse momentum representation and implemented it in the {\tt BFKLex} Monte Carlo event generator. This offers us the possibility to study its consequences in the global resummation program at a very exclusive level. We have found that its net effect is to improve the convergence in collinear regions, thus stabilizing the perturbative expansion even beyond the original multi- and quasi-multi-regge kinematics. The new collinear contributions take the form of a Bessel function of the first kind which reduces the total growth with energy of BFKL cross sections. Taking advantage of the event generator, we have found that the spread in transverse momentum of the typical scales within the BFKL ladder is greatly reduced by the collinear improvements. Quite remarkably, we have shown how the mean mini-jet multiplicity is also reduced when these higher order corrections are taken into account. It is now mandatory to further investigate how this improved resummation program affects exclusive observables at the Large Hadron Collider.

\begin{flushleft}
{\bf \large Acknowledgements}
\end{flushleft}
G.C. acknowledges support from the MICINN, Spain, under contract FPA2013-44773-P. 
A.S.V. acknowledges support from Spanish Government (MICINN (FPA2010-17747,FPA2012-32828)) and to the Spanish MINECO Centro de Excelencia Severo Ochoa Programme (SEV-2012-0249).

\end{document}